\documentclass[preprint,showpacs,preprintnumbers,amsmath,amssymb,superscriptaddress]{revtex4}
\usepackage{graphicx, epsfig}
\begin{document}

\preprint{ }

\title{Evidence for a correlated insulator to antiferromagnetic metal transition in CrN}

\author{P. A. Bhobe}
\email[corresponding author:]{preeti@spring8.or.jp}
\affiliation{Institute for Solid State Physics, The University of Tokyo, Kashiwa, Chiba 277-8581, Japan}
\affiliation{RIKEN, SPring-8 Centre, Sayo-cho, Sayo-gun, Hyogo 679-5148, Japan}
\author{A. Chainani}
\affiliation{RIKEN, SPring-8 Centre, Sayo-cho, Sayo-gun, Hyogo 679-5148, Japan}
\author{M. Taguchi}
\affiliation{RIKEN, SPring-8 Centre, Sayo-cho, Sayo-gun, Hyogo 679-5148, Japan}
\author{T. Takeuchi}
\affiliation{JASRI/SPring-8, Sayo-cho, Sayo-gun, Hyogo 679-5198, Japan}
\author{R. Eguchi}
\affiliation{Institute for Solid State Physics, The University of Tokyo, Kashiwa, Chiba 277-8581, Japan}
\affiliation{RIKEN, SPring-8 Centre, Sayo-cho, Sayo-gun, Hyogo 679-5148, Japan}
\author{M. Matsunami}
\affiliation{Institute for Solid State Physics, The University of Tokyo, Kashiwa, Chiba 277-8581, Japan}
\affiliation{RIKEN, SPring-8 Centre, Sayo-cho, Sayo-gun, Hyogo 679-5148, Japan}
\author{K. Ishizaka}
\affiliation{Institute for Solid State Physics, The University of Tokyo, Kashiwa, Chiba 277-8581, Japan}
\author{Y. Takata}
\affiliation{RIKEN, SPring-8 Centre, Sayo-cho, Sayo-gun, Hyogo 679-5148, Japan}
\author{M. Oura}
\affiliation{RIKEN, SPring-8 Centre, Sayo-cho, Sayo-gun, Hyogo 679-5148, Japan}
\author{Y. Senba}
\affiliation{JASRI/SPring-8, Sayo-cho, Sayo-gun, Hyogo 679-5198, Japan}
\author{H. Ohashi}
\affiliation{JASRI/SPring-8, Sayo-cho, Sayo-gun, Hyogo 679-5198, Japan}
\author{Y. Nishino}
\affiliation{RIKEN, SPring-8 Centre, Sayo-cho, Sayo-gun, Hyogo 679-5148, Japan}
\author{M. Yabashi}
\affiliation{RIKEN, SPring-8 Centre, Sayo-cho, Sayo-gun, Hyogo 679-5148, Japan}
\author{K. Tamasaku}
\affiliation{RIKEN, SPring-8 Centre, Sayo-cho, Sayo-gun, Hyogo 679-5148, Japan}
\author{T. Ishikawa}
\affiliation{RIKEN, SPring-8 Centre, Sayo-cho, Sayo-gun, Hyogo 679-5148, Japan}
\author{K. Takenaka}
\affiliation{Department of Crystalline Materials Science, Nagoya University, Nagoya 464-8603, Japan}
\affiliation{RIKEN, The Institute for Physical and Chemical Research, Wako, Saitama 351-0198, Japan}
\author{H. Takagi}
\affiliation{RIKEN, The Institute for Physical and Chemical Research, Wako, Saitama 351-0198, Japan}
\author{S. Shin}
\affiliation{Institute for Solid State Physics, The University of Tokyo, Kashiwa, Chiba 277-8581, Japan}
\affiliation{RIKEN, SPring-8 Centre, Sayo-cho, Sayo-gun, Hyogo 679-5148, Japan}
\date{\today}

\begin{abstract}
We investigate the electronic structure of Chromium Nitride (CrN) across the first-order magneto-structural transition at T$_N \sim$ 286 K. Resonant photoemission spectroscopy shows a gap in the $3d$ partial density of states at the Fermi level and an {\it On-site Coulomb energy U} $\sim$ 4.5 eV, indicating strong electron-electron correlations. Bulk-sensitive high resolution (6 meV) laser photoemission reveals a clear Fermi edge indicating an antiferromagnetic metal below T$_N$. Hard x-ray Cr $2p$ core-level spectra show T-dependent changes across T$_N$ which originate from screening due to coherent states as substantiated by cluster model calculations using the experimentally observed {\it U}. The electrical resistivity confirms an insulator above T$_N$ (E$_g \sim$ 70 meV) which becomes a disordered metal below T$_N$. The results indicate CrN transforms from a correlated insulator to an antiferromagnetic metal, coupled to the magneto-structural transition.
\end{abstract}
\maketitle

The competition between localization due to on-site Coulomb correlations between electrons and the kinetic energy gain due to band formation lies at the heart of diverse electronic and magnetic properties of transition metal oxides. It often leads to the so-called Mott-Hubbard metal-insulator transition which forms the basis for understanding unconventional phenomena, such as high-temperature superconductivity, colossal magneto-resistance, etc. Nitrides of transition metals also display a rich variety of interesting properties in accord with its oxide counterpart \cite{niwa}. Besides the high melting points, ultra hardness and corrosion resistance that generate tremendous technological interest, their fundamental properties also include superconductivity in VN, NbN with a T$_c$ as high as $\sim$16 K \cite{prat}, transition from a superconductor to a {\it Cooper-pair insulator} in TiN \cite{batu}, a magnetostructural transition as a function of temperature in CrN \cite{cor, brow, tsu, cons, quin}, superhard PtN$_2$, IrN$_2$ and OsN$_2$ \cite{crow, yon}, prediction of a displacive semiconductor-metal transition in IrN$_2$\cite{yu}, giant negative thermal expansion in an antiperovskite nitride Mn$_3$(Cu$_{1-x}$Ge$_x$)N \cite{take} etc.

CrN indeed seems to be quite peculiar as it does not show superconductivity like its isostructural near neighbours (TiN, VN, NbN, etc.). CrN is not even a metal at room temperature, but a cubic {\it Fm3m} paramagnetic insulator and undergoes a first order magneto-structural transition with hysterisis, to an orthorhombic{\it Pnma} antiferromagnet (AFM) below a Neel temp of T$_N \sim$ 286 K\cite{cor, brow, tsu, cons, quin}. The AFM ordering, as identified by neutron scattering\cite{cor}, consists of ferromagnetic (FM) (110) layers, stacked antiferromagnetically along the [110] direction, with a magnetic moment of 2.36 $\mu_B$ per Cr site. Such a magnetic structure is unique in itself and based on LSDA (local spin-density approximation) band structure calculations, it was shown to be responsible for the structural distortion to the {\it Pnma} orthorhombic phase \cite{filli-l}. In contrast to the general agreement on the magnetic ordering and crystal structure across T$_N$, the reports on its electronic properties are quite contradictory, with (i) metal to metal\cite{brow, tsu}, (ii) insulator to insulator\cite{sub, gall} and also, (iii) insulator to metal\cite{cons} transition across T$_N$. Theoretically, LSDA calculations\cite{filli-l, filli-b} could account for the magnetostructural transition, but predicts a metallic ground state for both phases. The results showed Cr $3d$ partial density of states (DOS) at the Fermi level (E$_F$), albeit with a depleted DOS for the AFM phase with a low carrier density of holes and electrons  (0.025 carriers of each sign), but no gap formation.  In contrast, LSDA+$U$ (where $U$ is the {\it On-site Coulomb energy} for Cr $3d$ electrons) \cite{adi} calculations for the AFM phase, with a $U$ of 3 eV, gives a direct gap in the DOS at E$_F$. A very recent study has identified a softening of bulk CrN under pressure due to a crossover from a localized to a molecular orbital electronic transition \cite{quin-nat}. Hence, the electronic structure of CrN is a controversial issue, and spectroscopic experiments of bulk CrN which exhibits the magnetostructural transition are urgently required to clarify the role of Cr $3d$ states for its electronic properties.  

With this aim, we have investigated the electronic structure of well-characterised polycrystalline samples of CrN. Resonant photoemission spectroscopy shows a gap in the $3d$ partial density of states and $U$$\sim$ 4.5 eV, indicating strong electron-electron correlations. However, high resolution ($\sim$6 meV) bulk-sensitive laser photoemission reveals a clear Fermi edge indicating an antiferromagnetic metal below T$_N$. A detailed analysis of the electrical resistivity confirms an insulator to metal transition across T$_N$. The results indicate CrN exhibits a coupled magneto-structural and insulator to metal transition.

\begin{figure}
\centering
\includegraphics[width=1\columnwidth]{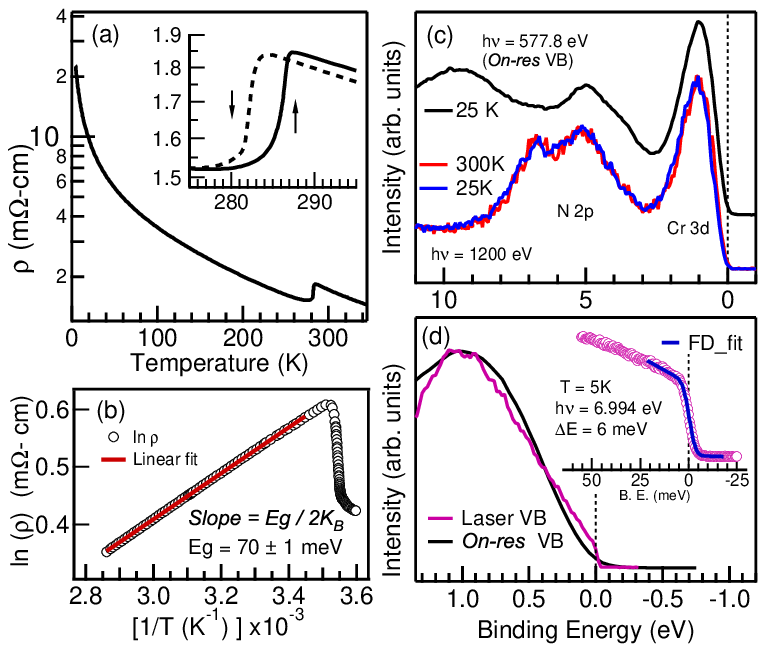}
\caption{\label{vb}(Colour online) (a) Temperature dependent resistivity, $\rho$(T), with inset showing the magnetostrucutral transition. (b) Using the expression $\rho(T) = \rho_0 e^{-(Eg/2k_BT)}$, a linear fit to ln$\rho$ Vs 1/T yeilds the band gap $E_g \sim$ 70 meV in the region above T$_N$. (c) VB spectra obtained using soft X-rays shows a highly reduced DOS at E$_F$. (d) Comparision of the On-resonance VB with Laser VB, showing a clear Fermi edge in the latter. Inset shows the Laser VB fitted with the resolution-convoluted Fermi Dirac function.} 
\end{figure}

CrN used in the present study was prepared by heating CrCl$_3$ (3N grade) in flowing NH$_3$ (5N grade) at 900$^{\circ}$C for 20 hours. This treated powder was ground, pressed into a pellet, and sintered at 1050$^{\circ}$C for 30 hours under gas flow of N$_2$ purified by filters (Nikka-Seiko, DC-A4 and GC-RX). The obtained CrN sample was found to be phase pure and was further characterized for it electrical transport and magnetic properties. The electrical resistivity as a function of temperature $\rho(T)$ is shown in Fig.\ref{vb}(a). The inset shows $\rho(T)$ on an expanded scale, indicating a clear hysterisis in the transition temperatures upon heating (T$_N =$ 286 K) and cooling (T$_N =$ 282 K). This is in very good agreement with a recent study, and indicates the high quality of the sample \cite{cons,quin}. Besides identifying the first order nature of the transition, $\rho(T)$ above T$_N$ shows an activation energy of 70 meV (Fig.\ref{vb}(b)), again in good agreement with samples showing a first order transition and confirms the insulating state above T$_N$ \cite{cons}. While $\rho(T)$ below T$_N$ also exhibits a negative temperature coefficient of resistivity (TCR), it does not follow an activated behavior. After exhausting all possibilities for an insulating behavior below T$_N$ (it does not fit variable range hopping in 2 or 3 dimensions), we have then checked for the M\"obius criterion which shows that d($ln\sigma$)/d($lnT$) $\to$ 0 as $T \to$ 0. This indicates a metal ground state, consistent with ref.\cite{cons}, inspite of the negative TCR. A detailed analysis is discussed in the supplementary text \cite{supp}. 

We then carried out extensive spectroscopy measurements of the valence band, Cr $2p$ and N $1s$ core-level photoemission (PES) and  X-ray absorption (XAS), as well as off- and on-resonant PES across the Cr ${2p \to 3d}$ excitation threshold. The soft/hard x-ray experiments were performed using synchrotron radiation at beamline BL17SU/BL29XU, SPring-8. Laser PES was performed using a Scienta R4000WAL electron analyzer and a vacuum-ultraviolet laser (h$\nu$= 6.994 eV)\cite{kiss}. T-dependent measurements were carried out using a liquid He flow-type cryostat. See \cite{supp} for details.

Fig.\ref{vb}(c) shows the valence band (VB) spectra of CrN measured with soft x-rays (h$\nu$ = 1200 eV and 577.8 eV), above and below T$_N$. The h$\nu$ = 1200 eV spectra show a high intensity peak  at 1.0 eV binding energy (BE) while broader features are observed in the region between 4 eV to 8 eV below $E_F$. The spectra show negligible change above and below T$_N$. The peak at 1 eV is dominated by Cr $3d$ states as it gets strongly enhanced in the resonant PES obtained with h$\nu$ = 577.8 eV, which corresponds to the main peak in the Cr $2p-3d$ XAS. The broad features between 4 - 8 eV are due to N $2p$ dominated states, in fair agreement with band structure calculations. The Cr $3d$ dominated peak shows negligible intensity at $E_F$, implying a gap in the Cr $3d$ partial density of states, in agreement with earlier results for the paramagnetic phase at 300 K i.e. above T$_N$ \cite{gall}. However, the absence of intensity below T$_N$ seems to contradict the ground state metal character inferred from the resistivity. In order to check the same with high-resolution($\sim 6 meV$) and bulk sensitivity($\lambda \sim100 \AA$), the VB spectra near $E_F$  were obtained using a Laser source in the AF phase, as shown in Fig.\ref{vb}(d). A clear difference is obtained in the bulk-sensitive spectrum with a sharp Fermi edge, confirming the metallic nature of CrN at low temperature(see inset). The existence of a metallic Fermi edge should have implications also for bulk-sensitive core-level spectra. At this point, such an interpretation seems a mere conjecture, but we have confirmed the observation of a T-dependent \textit{well-screened} feature in the Cr $2p$ core-level spectrum as discussed later. We first discuss the details of the valence band spectra in comparison to the known LSDA and LSDA + U results. While a finite DOS at $E_F$ corroborates well with the LSDA calculations\cite{filli-b}, it is important to note that the Cr $3d$ dominated peak occurs at 1.0 eV, whereas LSDA obtains the same feature at 0.5 eV BE. Further, in the AF phase LSDA yields a total DOS at E$_F$ to be about one-third the value for the 0.5 eV peak in the cubic phase. While, as is evident from the experimental data, the relative intensity at E$_F$ is far weaker (about one-tenth of the peak at 1.0 eV, normalizing the experimental and LSDA results to the Cr $3d$ 1.0 eV peak). Thus, though we find a metallic Fermi edge for the AF phase, the position of the Cr dominated peak and the highly reduced DOS observed experimentally, suggest additional effects beyond LSDA. This led us to the importance of strong electron-electron correlations as an important ingredient in the electronic structure of CrN. Consequently, we compare the experimental results with recent LSDA + U results \cite{adi} and find that the Cr $3d$ peak is correctly obtained at about 1 eV BE. However, for relevant $U$ values expected of chromium compounds, typically $U$ = 4 eV, the LSDA + U results give a band gap\cite{adi}. Thus it seems, LSDA + U could be valid for the insulating phase of CrN above T$_N$. However, to validate this picture, it is important to know the experimental $U$ for CrN. As is known from experiments on Cr metal \cite{huf}, $U$ can be determined by measuring the resonant photoemission(res-PES) across the Cr $2p-3d$ threshold and we have carried out the same for CrN. 

\begin{figure}
\centering
\includegraphics[width=1\columnwidth]{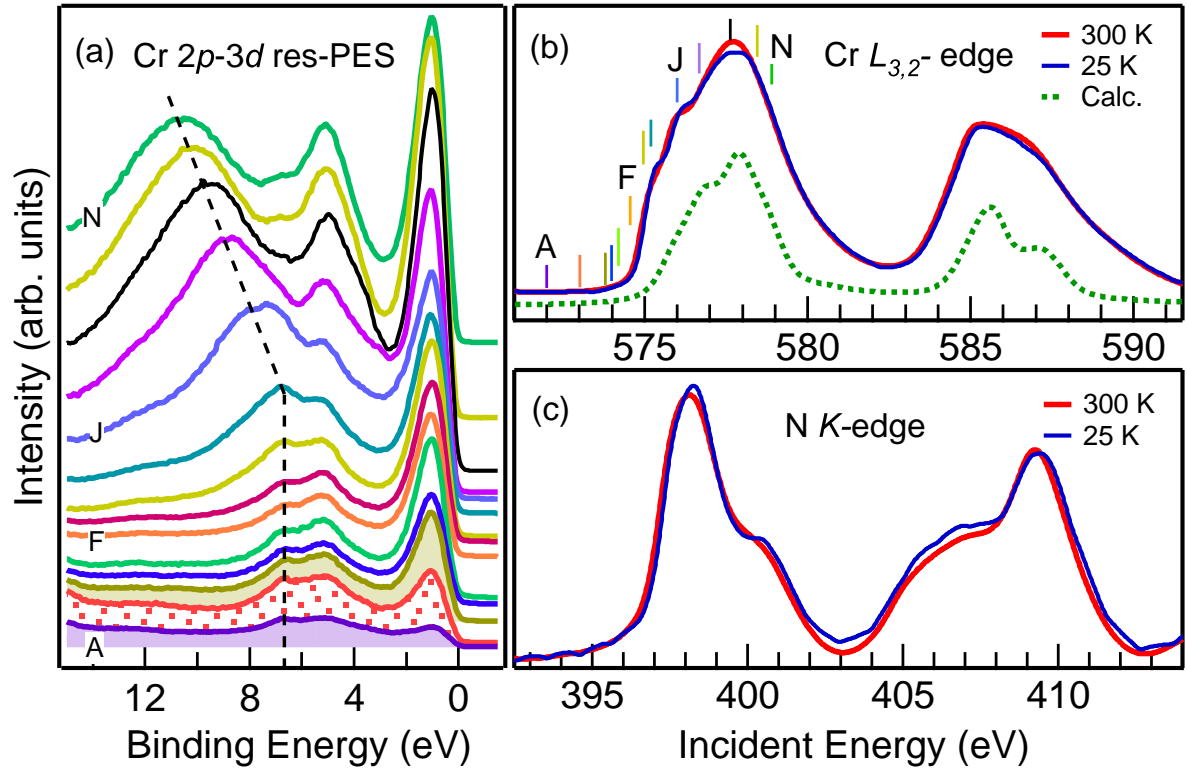}
\caption{\label{rpes}(Colour online)(a) Res-PES measured across the Cr $2p_{3/2} - 3d$ threshold at 25 K using incident photon energies indicated by tick marks on the Cr $L_{3,2}$ XAS curve shown in (b). The calculated XAS using the $U_{dd}$ obtained from res-PES compared to the experimental Cr $L_{3,2}$ XAS is also shown. (c) N K-edge XAS spectrum recorded at 300 K and 25 K.}
\end{figure}

Figure \ref{rpes} shows Cr $2p_{3/2} - 3d$ res-PES spectra for CrN measured at 25 K. The incident photon energy was varied from 572 to 579 eV, as indicated by tick marks on the Cr XAS curve in Fig.\ref{rpes}(b). We also measured res-PES in the high temperature phase (not shown here) and did not find any temperature dependent change. As seen in Fig.\ref{rpes}(a), strong enhancement of the PE intensity is found at 1.0 eV BE peak and around 6.5 eV BE as $h\nu$ is tuned across the Cr $2p_{3/2}$ XAS threshold. The intensity enhancement is due to the increase of photoemission cross section of the Cr $3d$ states. 
At higher $h\nu$, a strong intensity Auger feature shows up at BE higher than that of the Cr $3d$ dominated states found around 6.5 eV. This feature tracks the increase in $h\nu$ as shown with dotted line in Fig.\ref{rpes}(a). The evolution of the resonant feature into the regular Auger feature is an indicator of strong correlations in  CrN as also for several other materials \cite{fano, aber}. By substituting the appropriate BE value of the Auger peak for the corresponding incident photon energy, an estimate of the On-site Coulomb energy can be obtained using the equation, $U_{dd} = E_{2p} - (h\nu - E_{LVV}) - 2\epsilon_{3d}$. We obtain an estimate of $\sim$ 4.5 eV for CrN. It is also important to note that the obtained value of $U$  remains unchanged for the high temperature phase as the res-PES spectrum recorded at 300 K is identical to that of 25 K. 

In addition to the XAS for Cr $L_{3,2}$ edge, Fig. \ref{rpes} depicts N K-edge spectra recorded at 300 K and 25 K. With the change in temperature, small changes are observed in the Cr and N XAS that reflect the change in hybridization between Cr-N due to the associated magneto-structural transition. The magnetic stress generated in the crystal structure is known to be relieved by orthorhombic distortions that stabilize the {\it Pnma} phase \cite{filli-l}. These distortions lead to stronger hybridization of Cr-Cr bonds at the expense of Cr-N bonds. This has been confirmed by {\it ab initio} LSDA + U calculations that indicate a  charge transfer from the Cr orbitals that form the covalent Cr-N bonds towards the Cr-Cr bonds, at the phase transition \cite{quin-nat}. It is also noted that the Cr-N bond distance remains nearly constant while the cubic Cr-Cr bond splits into two non-equivalent bonds in the AF phase. Thus, we believe the small but finite changes in the N K-edge and Cr L-edge XAS reflects the related changes in the structure and hybridization.

\begin{figure}
\centering
\includegraphics[width=1\columnwidth]{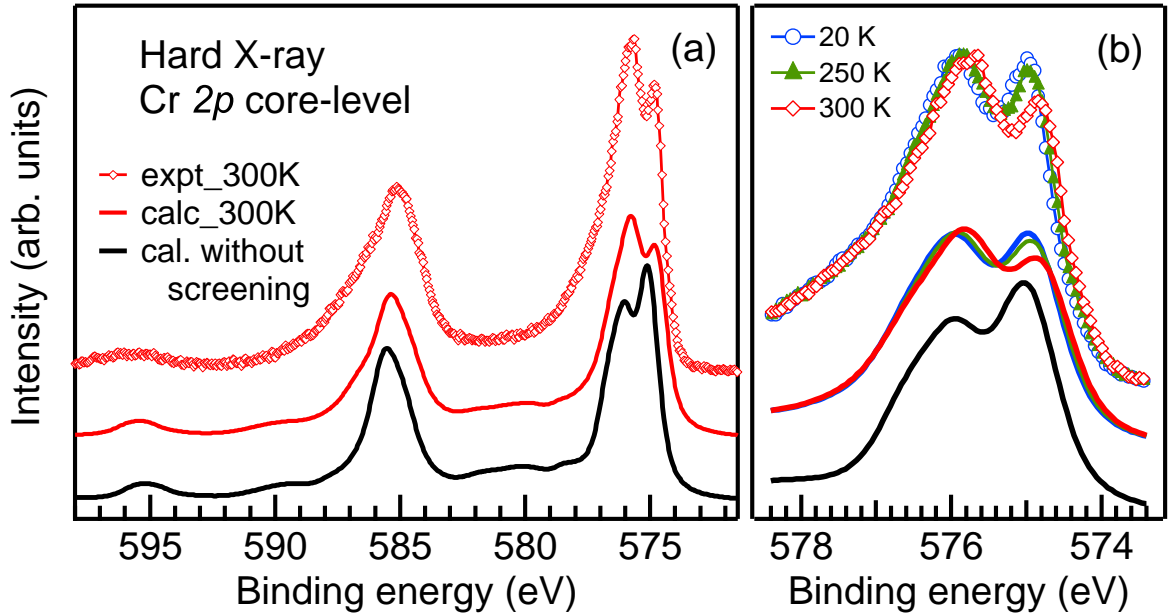}
\caption{\label{2pcore}(Colour online) (a) Cr $2p$ core-level spectra measured across T$_N$. The well-screened feature at the low BE side of $2p_{3/2}$ peak is reproduced in the calculated spectra using metallic screening channel at E$_F$, while the spectrum calculated without metallic screening does not match with the experimental spectra. (b) The $2p_{3/2}$ region shown on an expanded scale.}
\end{figure}

At the outset, having established strong correlations with a sizable $U$, it should reflect in a depletion of the DOS at $E_F$ which can eventually result in the opening of a gap. However, strong evidence of T-dependent metallicity is also seen in the Hard X-ray Cr $2p$ core-level PES as presented in Fig.\ref{2pcore}. The Cr $2p$ core-level shows a clear T-dependence across T$_N$ for the well-screened feature observed at 575 eV BE. The $Cr 2p_{3/2}$ main peak is observed at 575.9 eV BE. Note that the significant change in intensity occurs between the T = 300 K and 250 K spectra and then the spectra do not change down to 20 K, thus indicative of changes only across T$_N$. Similar well-screened features at the lower BE side of the main peak in core-level PES has been reported in several transition metal oxides with a direct relation to a metal-insulator transition, either as a function of T or doping\cite{chai, egu2}. In order to confirm this, we carried out cluster model calculations including a full multiplet structure and a screening channel derived from the states at $E_F$ responsible for metallicity. In particular, we used the experimentally observed $U$ = 4.5 eV, and a CrN$_{6}$ cluster, and the details are described in the supplementary text \cite{supp}. As is evident from the figure, the calculated spectra with screening channel shows good agreement with the data. With the same parameters, we also calculate the Cr XAS in agreement with experimental spectrum shown in Fig.\ref{rpes}(b). 

It was shown that for Cr based sulphides and selenides with a local magnetic moment of 2.9 - 3.0 $\mu_B$, the $Cr 2p_{3/2}$ core-level PES exhibited a splitting of 0.95 - 1.0 eV. Accordingly, the splitting of $\Delta E$ = 0.9 eV seen in our measurements below T$_N$ agrees well with the reported local magnetic moment of 2.36 $\mu_B$ per Cr site from neutron diffraction studies \cite{cor}. Qualitatively, the calculated spectrum without the inclusion of screening states should hence reflect the experimental spectrum at 300K. As is evident from Fig.\ref{2pcore}, such a calculated spectrum however shows a mismatch of intensities of the peaks at 575 and 575.9 eV. This can be understood if allowance is made for the 300 K thermal broadening at $E_F$($4K{_B}T \sim 100$ meV) as actually reflecting in the spectroscopic measurements, which overcomes the gap of 70 meV\cite{supp}.

The present experimental data show strong electron-electron correlations, validating the LSDA+U results for the cubic phase of CrN \cite{adi,quin-nat} which predict the opening of a direct gap even for $U$ as small as 3 eV. Though the predicted value of the gap $\sim$ 0.7 eV (for $U$ = 4 eV), is significantly larger than the experimental value of 70 meV, it indicates CrN is a correlation induced insulator above T$_N$. Additionally, it may be recalled that the observed On-site Coulomb energy $U$ remains unchanged across T$_N$, the Cr-Cr hybridization strength is expected to increase in the low temperature phase \cite{quin-nat} thereby increasing the effective bandwidth. This justifies our modeling of the metal phase with states at $E_F$ representing the coherent states in dynamical mean-field theory calculations of the Mott-Hubbard transition. In the same context, studies which report a semiconducting gap at low temperature in CrN also report an absence of the magneto-structural transition. An important distinction between a `conventional' Mott-Hubbard system and CrN is noted : instead of a paramagnetic metal to antiferromagnetic insulator, CrN exhibits a paramagnetic insulator to  antiferromagnetic metal phase transition.

In conclusion, we have performed temperature dependent Cr $2p$ core level, Nitrogen K- and Chromium L$_{3,2}$-edge absorption, off- and on-resonant photoelectron spectroscopy of CrN. Evidence for strong electron-electron correlations with an On-site Coulomb energy $U\sim$ 4.5 eV is obtained. The valence band and core-levels show evidence for a correlated insulator to metal transition across T$_N$. Model cluster calculations further establish that the T-dependent well-screened feature in the Cr $2p_{3/2}$ core-level spectrum is a signature of bulk screening from the coherent states of the highly correlated metallic antiferromagnet, CrN. 

P. A. Bhobe acknowledges the Japan Society for Promotion of Science for a postdoctoral fellowship.

\end{document}